\documentclass[pss,fleqn]{w-art}
\usepackage{times}
\usepackage{w-thm}
\usepackage[]{graphicx}
\begin{document}
\DOIsuffix{theDOIsuffix}
\Volume{XX}
\Issue{1}
\Month{01}
\Year{2003}
\pagespan{3}{}
\Receiveddate{\today}
\keywords{Hubbard model, Coulomb interaction, gauge theory}
\subjclass[pacs]{74.20.Fg, 74.72.h, 71.10.Pm}



\title[Spin-charge rotating local reference frames]{Spin-charge rotating local reference frames:
a unified U(2)=U(1)$\otimes$SU(2) approach to the  interacting electrons}


\author[T. K. Kopec]{T.K.Kopec\footnote{Corresponding
     author: e-mail: {\sf kopec@int.pan.wroc.pl}
     }\inst{1}} \address[\inst{1}]{Institute for Low Temperature and Structure Research, Polish Academy of Sciences,\\
POB 1410, 50-950 Wroclaw 2, Poland}
\begin{abstract}
A  spin-charge  unifying description  for the Hubbard model based on the time dependent local
gauge transformations is developed. The collective variables for charge and spin  are isolated in the form of the
space-time fluctuating U(1) phase field and rotating spin quantization axis governed
by the SU(2) symmetry, respectively. As a result interacting  electrons  appear as a composite 
objects consisting of bare fermions  with attached U(1) and SU(2) gauge fields.
We elaborate on the microscopic origins of the effective action with the Coulomb interaction 
that contain topological theta terms. Furthermore, we unravel the link between 
non-trivial multiply-connected topological structure of the U(2)=U(1)$\otimes$SU(2) configurational
 space for  gauge fields and  the instanton contribution to the statistical sum.
\end{abstract}
\maketitle                   





\section{Introduction}

Among the electronic Hamiltonians  relevant for interacting systems the  Hubbard model 
\cite{hubmod1,hubmod2,hubmod3} is considered as one  that contains  the essential ingredients for understanding the physics of correlated electrons.  In a many-body system the relevant physics
is encoded in the symmetries  and  in the Hubbard model they are represented 
by the charge U(1)  gauge and  spin  rotational SU(2) groups relevant for the occurrence
of the superconducting and magnetic orderings.
Interestingly, the  resulting U(2)=U(1)$\otimes$SU(2) symmetry group has a nontrivial
multiply connected   topological structure.
For multiply connected configuration spaces 
novel features can arise as    for  example in  the Aharonov-B\"ohm effect \cite{ab}
governed by  the multiply connected U(1) group manifold.
Since the homotopy class $\pi_1$[U(2)]$={Z}$ also forms a set of integer winding
 numbers	the topological structure of the configuration space is nontrivial, ambiguities may
arise when attempts are made to specify a value for the phase of a wave function for the
whole configuration space \cite{morandi}. 
Thus the problem we are facing is that of many-body
quantum mechanics on  a multiply connected configuration space.
In the present work we  develop a  spin-charge  unifying 
description  for interacting electrons given by the Hubbard model. It is based on the  time 
dependent local U(2) gauge transformations to disentangle  the Coulomb interaction. The collective
 variables for charge and spin are isolated in a form of the space-time fluctuating U(1) phase
 field and the rotating spin quantization axis governed by the SU(2) symmetry, respectively.
 As a result interacting  electrons  appear as  composite objects consisting of bare
fermions  with attached U(1) and SU(2) gauge fields and due nontrivial topology of the U(2) group an
instanton contribution to the statistical sum emerges.

%
\section{Hubbard Hamiltonian}
%
Our starting point is the purely fermionic Hubbard Hamiltonian
 ${\cal H}\equiv {\cal H}_t+{\cal H}_U$:
\begin{eqnarray}
{\cal H}=
 -t\sum_{\langle {\bf r}{\bf r}'\rangle,\alpha}
 [c^{\dagger }_{{\alpha}}({\bf r})
c_{\alpha }({\bf r}')+{\rm h.c.}]
+
U\sum_{\bf r}n_{\uparrow} ({\bf r}) n_{\downarrow}({\bf r}).
\label{mainham}
\end{eqnarray}
Here, $\langle {\bf r},{\bf r}'\rangle$  runs
over the nearest-neighbor (n.n.) sites, $t$  is the  hopping amplitude, $U$ stands for
the Coulomb repulsion,
while the operator $c_{\alpha }^\dagger({\bf r})$
creates an electron with spin $\alpha=\uparrow ,\downarrow$ at the lattice site ${\bf r}$,
 where  ${n}_{\alpha}({{\bf r}})= c^\dagger_{\alpha}({\bf r})
c_{\alpha}({\bf r})$. Usually, working in the grand canonical ensemble  a term
is added to ${\cal H}$ in Eq.(\ref{mainham})to control the average number of electrons,
${\cal H}\to {\cal H}-\mu\sum_{\bf r}{n}({\bf r})$
with $\mu$  being  the chemical potential and
${n}({{\bf r}})= n_{\uparrow} ({\bf r})+n_{\downarrow}({\bf r})$
 the  number operator.
It is customary to introduce Grassmann fields,
$c_\alpha({\bf r}\tau)$ 
depending on the ``imaginary time" $0\le\tau\le \beta\equiv 1/k_BT$,
(with $T$ being the temperature)
that satisfy the anti--periodic condition
$c_{\alpha}({\bf r}\tau)=-c_{\alpha}({\bf r}\tau+\beta)$,
to write the path integral for the statistical sum
${\cal Z}=\int\left[{\cal D}\bar{c}  {\cal D}{c}
\right]e^{-{\cal S}[\bar{c},c]}$
with the fermionic action 
\begin{eqnarray}
{\cal S}[\bar{c},c]={\cal S}_B[\bar{c},c]+\int_0^\beta d\tau{\cal H}[\bar{c},c],
 \end{eqnarray}
that contains the fermionic  Berry term \cite{berry}:
${\cal S}_B[\bar{c},c]=\sum_{{\bf r}\alpha }\int_0^\beta d\tau
 \bar{c}_{\alpha }({\bf r}\tau)\partial_\tau{c}_{\alpha }({\bf r}\tau)$
that will play an important role in our considerations.
%
\section{Spin-charge U(2) reference frames}

For strongly correlated system it is  crucial to construct a covariant formulation of the theory
which  naturally  preserves  the  spin-rotational symmetry present in the Hubbard Hamiltonian.
In order to maintain spin rotational invariance, one
should consider the spin-quantization axis to be a priori
arbitrary and integrate over all possible directions in
the partition function.
For this purpose the density--density product 
in Eq.(\ref{mainham}) we write, following Ref.\cite{schulz}, in
a spin-rotational invariant way:
\begin{equation}
{\cal H}_U=U\sum_{{\bf r} }\left\{\frac{1}{4}{n }^2({{\bf r}}\tau)
-\left[{\bf \Omega} ({\bf r}\tau)\cdot{\bf S} ({\bf r}\tau)\right]^2\right\},
\label{huu}
\end{equation}
where
 $S^a({\bf r}\tau)=\frac{1}{2}\sum_{\alpha\alpha'}c^\dagger_{\alpha}({\bf r}\tau)
\hat{\sigma}_{\alpha\alpha'}^a c_{\alpha'}({\bf r}\tau)$
denotes the vector spin operator ($a=x,y,z$) with $\hat{\sigma}^a$ being
the Pauli matrices. The unit vector 
 ${\bf \Omega} ({\bf r}\tau)=[\sin\vartheta({\bf r}\tau)\cos\varphi({\bf r}\tau),
\sin\vartheta({\bf r}\tau)\sin\varphi({\bf r}\tau),
\cos\vartheta({\bf r}\tau)]$
written in terms of polar angles labels
varying in space-time  spin quantization axis. 
The spin--rotation invariance is made explicit
by performing the angular integration over  ${\bf\Omega}({\bf r}\tau)$
at each site and time. By decoupling spin and charge density terms 
in Eq.(\ref{huu}) using auxiliary fields
$\varrho({\bf r}\tau)$ and $iV({\bf r}\tau)$ respectively,
we write down the partition function in the form \cite{popov}
\begin{eqnarray}
{\cal Z}&=&\int[{\cal D}{\bf \Omega}]\int[{\cal D} V{\cal D}{\varrho}
 ]\int\left[
  {\cal D}\bar{c}{\cal D}c
\right]
e^{-{\cal S}\left[{\bf \Omega},V,{\varrho},\bar{c},c\right]}.
\label{zfun}
\end{eqnarray}
where $[{\cal D}{\bf \Omega}]\equiv
\prod_{{\bf r}\tau_k}
\frac{\sin\vartheta({\bf r}\tau_k)d\vartheta ({\bf r}\tau_k)d\varphi ({\bf r}\tau_k)}{4\pi}$
is the spin-angular integration measure.
The  effective action reads:
\begin{eqnarray}
{\cal S}\left[{\bf \Omega},V,{\varrho},\bar{c},c\right]&=&
\sum_{ {\bf r} }\int_0^\beta
 d\tau
\left[\frac{{\varrho}^2 ({\bf r}\tau)}{U}+\frac{V^2 ({\bf r}\tau)}{U}\right.
+\left.iV ({\bf r}\tau)n ({\bf r}\tau)
+2{\varrho} ({\bf r}\tau){\bf \Omega} ({\bf r}\tau)\cdot {\bf S} ({\bf r}\tau)\right]
\nonumber\\
&+&{\cal S}_B[\bar{c},c]+\int_0^\beta d\tau{\cal H}_t[\bar{c},c].
\label{sa}
\end{eqnarray}
Simple Hartree-Fock  theory won't work for a Hubbard model in which $U$
 is the largest energy in the problem.
One has to isolate strongly fluctuating modes generated by the Hubbard term  according to the
charge U(1) and spin SU(2) symmetries.
To this end  we write the fluctuating ``imaginary chemical potential" $iV ({\bf r}\tau)$ as a sum of
a static $V_{0 }({\bf r})$ and periodic function
$V({\bf r}\tau)=V_0({\bf r})+\tilde{V}({\bf r}\tau)$
using  Fourier series
$\tilde{V}({\bf r}\tau)=\frac{1}{\beta}	\sum_{n=1}^\infty
[\tilde{V}({\bf r}\omega_n)e^{i\omega_n\tau}+c.c.]$
with $\omega_n=2\pi n/\beta$ ($n=0,\pm1,\pm2$)
being the (Bose) Matsubara frequencies.
Now, we introduce the U(1) {\it phase } field ${\phi} ({\bf r}\tau)$
via the Faraday--type relation \cite{kopec}
\begin{equation}
\dot{\phi} ({\bf r}\tau)\equiv\frac{\partial\phi ({\bf r}\tau)}
{\partial\tau}=e^{-i\phi ({\bf r}\tau)}\frac{1}{i}
\frac{\partial}{\partial\tau} e^{i\phi ({\bf r}\tau)}=\tilde{V} ({\bf r}\tau).
\label{jos}
\end{equation}
Furthermore, by performing the local gauge transformation to the {\it new} fermionic
variables $f_{\alpha}({\bf r}\tau)$:
\begin{eqnarray}
\left[\begin{array}{c}
c_{\alpha }({\bf r}\tau)\\
\bar{c}_{\alpha }({\bf r}\tau)
\end{array}\right]=
\left[\begin{array}{cc}
z({\bf r}\tau)&0\\
0& \bar{z}({\bf r}\tau)
\end{array}\right]
\left[\begin{array}{c}
f_{\alpha }({\bf r}\tau)\\
\bar{f}_{\alpha }({\bf r}\tau)
\end{array}\right]
\label{sing1}
\end{eqnarray}
where the unimodular parameter  $|z({\bf r}\tau)|^2=1$ satisfies $z({\bf r}\tau)=e^{i\phi ({\bf r}\tau)}$,
we remove the imaginary term $i\int_0^\beta d\tau\tilde{V}({\bf r}\tau)n({\bf r}\tau)$
for all the Fourier modes
of the $V ({\bf r}\tau)$ field, except for  the zero frequency.
Subsequent SU(2) transformation from $f_{\alpha}({\bf r}\tau)$ to  $h_{\alpha}({\bf r}\tau)$
operators,
\begin{eqnarray}
\left[\begin{array}{c}
f_{\uparrow }({\bf r}\tau)\\
{f}_{\downarrow}({\bf r}\tau)
\end{array}\right]=
\left[
\begin{array}{cc}
\zeta_{ \uparrow }({\bf r}\tau) & -\bar{\zeta}_{\downarrow}({\bf r}\tau) \\
\zeta_{\downarrow}({\bf r}\tau) & \bar{\zeta}_{\uparrow }({\bf r}\tau)
\end{array}
\right]\left[\begin{array}{c}
h_{1 }({\bf r}\tau)\\
{h}_{\downarrow}({\bf r}\tau)
\end{array}\right],\quad {\bf R}({\bf r}\tau) \equiv \left[
\begin{array}{cc}
\zeta_{ \uparrow }({\bf r}\tau) & -\bar{\zeta}_{\downarrow}({\bf r}\tau) \\
\zeta_{\downarrow}({\bf r}\tau) & \bar{\zeta}_{\uparrow }({\bf r}\tau)
\end{array}
\right]
\label{sing2}
\end{eqnarray}
with the constraint
$|\zeta_{\uparrow }({\bf r}\tau)|^2 +|\zeta_{\downarrow}({\bf r}\tau)|^2=1$ takes away the
rotational dependence on ${\bf \Omega}({\bf r}\tau)$ in the spin sector.
This is done by means  of the Hopf map
${\bf R}({\bf r}\tau) \hat{\sigma}^z{\bf R}^\dagger({\bf r}\tau)
 =\hat{{\sigma}}\cdot{\bf \Omega}({\bf r}\tau)$
that is based on the enlargement from two-sphere $S_2$ to the three-sphere $S_3\sim SU(2)$.
The unimodular constraint
can be resolved by using the parametrization
\begin{eqnarray}
\zeta_{\uparrow}({\bf r}\tau)& = & e^{-\frac{i}{2}[\varphi({\bf r}\tau)+\chi({\bf r}\tau)]}
\cos\left[\frac{\vartheta({\bf r}\tau)}{2}\right]
\nonumber\\
\zeta_{\downarrow}({\bf r}\tau)&=&e^{\frac{i}{2}[\varphi({\bf r}\tau)-\chi({\bf r}\tau)]}
\sin\left[\frac{\vartheta({\bf r}\tau)}{2}\right]
\label{cp1}
\end{eqnarray}
with the  Euler
angular variables $\varphi({\bf r}\tau),\vartheta({\bf r}\tau)$ and $\chi({\bf r}\tau)$, respectively.
Here, the  extra variable $\chi({\bf r}\tau)$   represents the U(1) gauge freedom  of the theory
as a consequence of $S_2\to S_3$ mapping. One can summarize Eqs (\ref{sing1}) and (\ref{sing2})
by the single joint gauge transformation exhibiting electron operator factorization
$c_{\alpha }({\bf r}\tau)=\sum_{\alpha'}
{ \mathcal U}_{\alpha\alpha'} ({\bf r}\tau)h_{\alpha'}({\bf r}\tau)$,
where 
${\mathcal U}({\bf r}\tau) =z({\bf r}\tau)
{\bf R} ({\bf r}\tau)$
is  a U(2) matrix  which
rotates the spin-quantization axis at site ${\bf r}$ and time $\tau$.
This  reflects the composite nature of the interacting electron formed from
bosonic spinorial and charge degrees of freedom given by  ${ R}_{\alpha\alpha'} ({\bf r}\tau)$
and $z({\bf r}\tau)$, respectively as well as  remaining fermionic part $h_{\alpha}({\bf r}\tau)$.
Accordingly, the integration measure over the group manifold becomes
\begin{eqnarray}
\int[{{\cal D}{\phi}}{{\cal D}{\bf \Omega}}]\equiv\sum_{\{m({\bf r})\}}\prod_{\bf r}\int_{0}^{2\pi}d\phi_0({\bf r})
\int d{\bf \Omega}_0({\bf r})
\int\limits^{\phi({\bf r}\beta)=\phi_0({\bf r})+2\pi m({\bf r})}_{\phi({\bf r}0)=\phi_0({\bf r})}
{\cal D}\phi({\bf r}\tau)\int\limits_{{\bf \Omega}({\bf r}0)={\bf \Omega}_0}^{{\bf \Omega}({\bf r}\beta)=
{\bf \Omega}_0} {\cal D}{\bf \Omega}({\bf r}\tau)
\label{measure}
\end{eqnarray}
where $\int d{\bf \Omega}\dots=\frac{1}{4\pi}\int_0^\pi\sin\theta d\vartheta\int_0^{2\pi} d\varphi\dots$
and $[{\cal D}{\bf \Omega}({\bf r}\tau)]=\prod_k d{\bf \Omega}({\bf r}\tau_k)$.
Here, $m\in {Z}$ labels equivalence classes of homotopically connected
paths \cite{schulman}. 
%
\section{Effective phase-angular action}

Anticipating that spatial and temporal fluctuations of the fields $V_0({\bf r})$
and $\varrho({\bf r }\tau)$ will be energetically penalised,
since they are gaped and decouple from the angular and phase variables.
Therefore, in order to make further progress towards
we next subject the functional  to a  saddle
point analysis. 
The expectation value of the static (zero frequency) part of
the fluctuating electrochemical potential $V_0(r)$ we calculate
by the saddle point method to give
$V_0(r)=i\left( \mu-\frac{U}{2}n_h \right)\equiv i\bar{\mu}$
where $\bar{\mu}$ is the chemical potential
with a Hartree shift originating from the saddle-point value of the static variable $V_0({\bf r})$ with $n_h=n_{h\uparrow}+n_{h\downarrow}$ and $n_{h\alpha}=
\langle\bar{h}_{\alpha }({\bf r}\tau)h_{\alpha }({\bf r}\tau)\rangle$.
Similarly in the magnetic sector
\begin{eqnarray}
\rho({\bf r}\tau)=\left\{
\begin{array}{r}
(-1)^{\bf r}\Delta_c\\
\pm\Delta_c
\end{array}
\right.
\label{spaff}
\end{eqnarray}
where $\Delta_c=U\langle S^z({\bf r}\tau \rangle$ sets the magnitude for the 
Mott-charge gap. The two choices delineated in Eq.(\ref{spaff}) correspond to the saddle point
of the  ``antifferomagnetic" (with staggering  $\Delta_c$) or ``ferromagnetic type".
Note that the notion ferromagnetic (antifferomagnetic) here does not mean
an actual  long--range ordering - for this the angular spin-quantization variables 
have to be ordered as well.
In the new variables the action in Eq.(\ref{sa}) assumes the form
${\cal S}\left[{\bf \Omega},\phi,{\varrho},\bar{h},h\right]=
{\cal S}_B[\bar{h},h]+\int_0^\beta d\tau{\cal H}_{\bf \Omega,\phi}[\rho,\bar{h},h] 
+{\cal S}_0\left[\phi\right]+2\sum_{\bf r }\int_0^\beta d\tau
{\bf A}({\bf r}\tau)\cdot {\bf S}_{h }({\bf r}\tau)$,
where ${\bf S}_{h}({\bf r}\tau)=\frac{1}{2}\sum_{\alpha\gamma}\bar{h}_{\alpha }({\bf r}\tau)
\hat{\sigma}_{\alpha\gamma} h_{\gamma}({\bf r}\tau)$. Furthermore,
\begin{eqnarray}
S_0[\phi]=\sum_{\bf r}\int_0^\beta d\tau\left[\frac{\dot{\phi}^2({\bf r}\tau)}{U} 
+\frac{1}{i}\frac{2\mu}{U}\dot{\phi}({\bf r}\tau) \right]
\label{sphi}
\end{eqnarray}
stands for the kinetic  and Berry term of the U(1) phase field in the charge sector.%
The SU(2) gauge transformation in Eq.(\ref{sing2}) and the  fermionic Berry term 
generate SU(2) potentials given  by
${\bf R} ^\dagger({\bf r}\tau){\partial}_{\tau}
{\bf R} ({\bf r}\tau) ={\bf R} ^\dagger\left(\dot{\varphi}\frac{\partial}{\partial\varphi}
+\dot{\vartheta}\frac{\partial}{\partial\vartheta}
+\dot{\chi}\frac{\partial}{\partial\chi} \right)
{\bf R} 
=   -
{\hat{\sigma}}\cdot {\bf A}({\bf r}\tau)$,
where
\begin{eqnarray}
A^x({\bf r}\tau)&=&\frac{i}{2}\dot{\vartheta}({\bf r}\tau)\sin\chi({\bf r}\tau)
-\frac{i}{2}\dot{\varphi}({\bf r}\tau)\sin\theta({\bf r}\tau)\cos\chi({\bf r}\tau)
\nonumber\\
A^y({\bf r}\tau)&=&\frac{i}{2}\dot{\vartheta}({\bf r}\tau)\cos\chi({\bf r}\tau)
+\frac{i}{2}\dot{\varphi}({\bf r}\tau)\sin\theta({\bf r}\tau)\sin\chi({\bf r}\tau)		
\nonumber\\
A^z({\bf r}\tau)&=&\frac{i}{2}\dot{\varphi}({\bf r}\tau)\cos\vartheta({\bf r}\tau)
+\frac{i}{2}\dot{\chi}({\bf r}\tau).
\end{eqnarray}
are the SU(2) gauge potentials.
The fermionic sector, in turn, is governed by the effective Hamiltonian
\begin{eqnarray}
 &&{\cal H}_{\bf \Omega,\phi} = 
	\sum_{{ \bf r}}{\varrho} ({\bf r}\tau) [\bar{h}_{{\uparrow} }({\bf r}\tau)h_{\uparrow  }({\bf r}\tau)-
\bar{h}_{{\downarrow} }({\bf r}\tau)h_{\downarrow  }({\bf r}\tau)]
\nonumber\\
&&-t\sum_{\langle {\bf r},{\bf r}'\rangle} \sum_{\alpha\gamma}
 \left[{\mathcal U}^\dagger ({\bf r}\tau){\mathcal U}_{ }({\bf r'}\tau)\right]_{\alpha\gamma}
\bar{h}_{{\alpha} }({\bf r}\tau)h_{\gamma }({\bf r}'\tau)
-\bar{\mu}\sum_{{ \bf r}\alpha}
\bar{h}_{\alpha }({\bf r}\tau)
 h_{\alpha }({\bf r}\tau),
\label{explicit}
\end{eqnarray}
The result  of the gauge transformations is that we have managed 
to cast the strongly correlated
problem into a system of mutually non-interacting fermions, submerged in the
bath of strongly fluctuating U(1) and SU(2)  fields  whose dynamics is governed  by the
energy scale set by the Coulomb interaction  $U$  coupled to fermions
via hopping term  and with the Zeeman-type contribution
with the massive field ${\varrho} ({\bf r}\tau)$.

In analogy to the charge U(1) field the SU(2)  spin system exhibit  emergent dynamics.
Integration of fermions  will generate the kinetic term
for the SU(2) rotors 
\begin{eqnarray}
{\cal S}_0[{\bf \Omega}]&=&-\frac{1}{2}\times 4\sum_{\bf rr'}\int_0^\beta  d \tau d\tau'
	\sum_{ab} A^a({\bf r}\tau) A^b({\bf r'}\tau') 
\nonumber\\
&\times& \sum_{\bf r'}\langle S^a_{h}({\bf r}\tau)
S^b_{h}({\bf r'}\tau')\rangle
-2\sum_{\bf rr'}\int_0^\beta  d \tau 
	{\bf A}({\bf r}\tau) \cdot \langle{\bf S}_h({\bf r'}\tau') \rangle
\end{eqnarray}
with
\begin{eqnarray}
\langle S^a_{h}({\bf r}\tau)
S^b_{h}({\bf r}'\tau')\rangle=-\frac{1}{4}\times 2\delta_{ab}\frac{1}{{\cal E}_s},
\quad
\langle S^z_h({\bf r}\tau)\rangle=\frac{\Delta_c}{U}
\end{eqnarray}
In the AF phase, at the half-filling, it assumes the
staggered form ${\varrho} ({\bf r}\tau)=\Delta_c (-1)^{\bf r}$ with $\Delta_c$ being the
charge gap $\Delta_c\sim U/2$ for $U/t\gg 1$,
with $E_{\bf k}=\sqrt{\varepsilon^2_{\bf k}+\Delta_c^2}$, ${\bf Q}=(\pi,\pi)$ and
$\varepsilon_{\bf k}=\epsilon_{\bf k}-{\bar{\mu}}$.
At zero temperature
$\lim_{T\to 0}\frac{1}{{\cal E}_s}=\frac{\Delta^2_c}{2E^3_{\bf k}}$
so that, in the limit $U/t\gg 1$ one obtains ${\cal E}_s\sim 2\Delta_c=U$.
However, a nonzero  value of $\Delta_c$ does not  imply the existence of anfiferromagnetic  long--range 
order. For this the angular degrees of freedom ${\bf \Omega}({\bf r}\tau)$ have also to be ordered,
whose  low-lying  excitations  are in the form of spin waves.
Therefore the kinetic term in the spin sector becomes
${\cal S}_0[{\bf \Omega}]=-\frac{1}{{\cal E}_s}\sum_{\bf r}
\int_0^\beta  d \tau  {\bf A}({\bf r}\tau)\cdot{\bf A}({\bf r}\tau)$
so that the total kinetic energy ${\cal S}[{\bf \Omega},\phi]
={\cal S}_0[{\phi}]+{\cal S}_0[{\bf \Omega}]$ for U(1) and SU(2) rotors becomes
\begin{eqnarray}
{\cal S}_0[{\bf \Omega},\phi]&=&
\sum_{\bf r}\int_0^\beta d\tau\left\{\frac{\dot{\phi}^2({\bf r}\tau)}{U} 
+\frac{1}{i}\frac{2\mu}{U}\dot{\phi}({\bf r}\tau) \right.
+
\frac{1}{4{\cal E}_s}
\left[{ \dot{\vartheta}^2({\bf r}\tau)+\dot{\varphi}^2({\bf r}\tau)+\dot{\chi}^2({\bf r}\tau)}
\right.
\nonumber\\
&+&
\left. {2}\dot{\varphi}({\bf r}\tau)
\dot{\chi}({\bf r}\tau)\cos{\vartheta}({\bf r}\tau)
\right]
+	\left.\frac{\Delta_c}{iU}
 \left[ \dot{\varphi}({\bf r}\tau)\cos\vartheta({\bf r}\tau)
+\dot{\chi}({\bf r}\tau) \right]\right\}.
\label{phiaction2}
\end{eqnarray}
The distinctive feature of Eq.(\ref{phiaction2}) is th  presence of the
geometric  Berry contributions that signify topological features of the
underlying field theory \cite{kopec1}.

\section{Summary}
%

In this work we have observed that the important symmetries  of the Hubbard model  given by
by the charge U(1)  gauge and  spin  rotational SU(2) groups imply that
the quantum dynamics is governed  non-trivially by 
the multiply connected U(2)=U(1)$\otimes$SU(2) group manifold. 
 Following this route
we have 	 developed a  spin-charge  unifying  description  for interacting electrons given by the Hubbard model. It is based on the  time 
dependent local gauge transformations to disentangle  the Coulomb interaction. The collective
variables for charge and spin are isolated in a form of the space-time fluctuating  phase
field and the rotating spin quantization axis.
As a result interacting  electrons  appear as a composite 
objects consisting of bare fermions  with attached  gauge fields.
Our  results  form  a general framework
tailored to describe eg. both magnetic  and superconducting order
which is rooted in microscopic of strongly correlated electrons
and their basic symmetries and interactions \cite{kopec2}.
In particular, it will enable to identify
the degrees of freedom of the hierarchy of low-energy effective
theories, and the way they are associated with
the global symmetries.

\begin{acknowledgement}
  This work was supported by the Ministry of Education and Science
(MEN) under Grant No. 1 P03B 103 30  in the years 2006--2008.
\end{acknowledgement}

\end{document}